\documentstyle[12pt]{article}
\begin{document}
\begin{titlepage}
\centerline{\bf HAMILTONIAN FORMULATION OF TELEPARALLEL}
\centerline{\bf THEORIES OF GRAVITY IN THE TIME GAUGE}
\vskip 1.0cm   
\bigskip
\centerline{\it J. W. Maluf$\,^{*}$ and A. A. Sousa}
\centerline{\it Departamento de F\'isica}
\centerline{\it Universidade de Bras\'ilia}
\centerline{\it C.P. 04385}
\centerline{\it 70.919-970  Bras\'ilia, DF}  
\centerline{\it Brazil}
\date{}
\begin{abstract}
We consider the most general class of teleparallel theories
of gravity quadratic in the torsion tensor, and carry out a
detailed investigation of its Hamiltonian formulation
in the time gauge. Such general class is given by a
three-parameter family of theories. A consistent
implementation of the Legendre transform reduces
the original theory to a one-parameter theory determined
in terms of {\it first class} constraints. The free
parameter is fixed by requiring the Newtonian limit. The
resulting theory is the teleparallel equivalent of general
relativity.

\end{abstract}
\thispagestyle{empty}
\vfill
\par
\noindent (*) e-mail: wadih@fis.unb.br
\end{titlepage}
\newpage

\noindent {\bf I. Introduction}\par
\bigskip
It is well known that Einstein's general relativity can be
obtained from several distinct Lagrangian formulations.
One suitable formulation is the teleparallel
equivalent of general relativity (TEGR), which is defined in
terms of tetrad fields $e^a\,_\mu$ ($ a,\,\mu $ are
SO(3,1) and space-time indices, respectively)
and actually represents an alternative
geometrical framework for Einstein's equations. The Lagrangian
density for the tetrad field in the TEGR is given by
a sum of quadratic terms
in the torsion tensor $T^a\,_{\mu \nu} =
\partial_\mu e^a\,_\nu - \partial_\nu e^a\,_\mu$, which is
related to the anti-symmetric part of the connection
$\Gamma^\lambda _{\mu \nu}= e^{a\lambda}\partial_\mu e_{a\nu}$.
The curvature tensor constructed out of the latter
vanishes identically. This connection defines a space with
teleparallelism, or absolute parallelism\cite{Schouten}.

In a space-time with an underlying tetrad field two  vectors at
distant points may be called parallel\cite{Moller1} if they have
identical components with respect to the local tetrads at
the points considered. Thus consider a vector field $V^\mu(x)$.
At the point $x^\lambda$ its tetrad components are given by
$V^a(x)=e^a\,_\mu(x)V^\mu(x)$. For the tetrad components
$V^a(x+dx)$ it is easy to show that $V^a(x+dx)=V^a(x)+DV^a(x)$,
where $DV^a(x)=e^a\,_\mu(\nabla_\lambda V^\mu)dx^\lambda$. 
The covariant derivative $\nabla$ is constructed out of the
connection
$\Gamma^\lambda_{\mu \nu}=e^{a\lambda}\partial_\mu e_{a\nu}$.
Therefore such connection defines a condition for absolute
parallelism in space-time. The tetrad fields are thus required
to transform under the global SO(3,1) group.

The Lagrangian density for the TEGR is
based on the relation

$$eR(e)\;=\; 
-e({1\over 4}T^{abc}T_{abc}\,
+\,{1\over 2}T^{abc}T_{bac}\,-\,T^aT_a)\,+\,
2\partial_\mu(eT^{\mu})\;,\eqno(1)$$

\noindent which can be verified by substituting the Levi-Civita
connection  $^0\omega_{\mu ab}$ into the scalar
curvature $R(e)$ on the left hand side of (1) by means of the
relation
$^0\omega_{\mu ab}\,=\,-K_{\mu ab}$, where
$K_{\mu ab}$ is the contorsion tensor:
$K_{\mu ab}={1\over 2}e_a\,^\lambda e_b\,^\nu(T_{\lambda \mu \nu}+
T_{\nu \lambda \mu}-T_{\mu \nu \lambda}$).

In empty space-time the Lagrangian density for the TEGR 
is given by\cite{Hehl1,Kop1,Nester,Maluf1,Maluf2}

$$L(e)\;=\;-ke({1\over 4}T^{abc}T_{abc}+{1\over 2} T^{abc}T_{bac}-
T^aT_a)\;,\eqno(2)$$

\noindent where $k={1\over {16\pi G}}$, $e=det(e^a\,_\mu)$ and
$T_a=T^b\,_{ba}$. As usual, tetrad fields convert space-time
into SO(3,1) indices and vice-versa. Let
${{\delta L}\over {\delta e^a\,_\mu}}$ denote the field equation
satisfied by $e^a\,_\mu$. It can be verified by explicit calculations
that the latter are equivalent to Einstein's equations in tetrad
form\cite{Maluf1}:

$${{\delta L}\over {\delta {e^{a\mu}}}}\;=\;
{1\over 2}e\lbrace R_{a\mu}(e)\,
-\,{1\over 2}e_{a\mu}R(e)\rbrace\;.\eqno(3)$$

This theory has been considered long ago by
M\o ller\cite{Moller1,Moller2}, although not precisely in the form
presented above. More recently it has been reconsidered  as a
gauge theory of the translation group\cite{Cho,Pereira}. Interesting
developments\cite{Mielke} have been achieved in the context
of Ashtekar variables\cite{Ashtekar}. It has been shown that in the
teleparallel geometry the (complex) Hamiltonian becomes quadratic
in the new field momenta, and that Einstein's equations become
formally of the Yang-Mills type.

An important feature of the above formulation is that the tetrad
fields $e^a\,_\mu$ transform under the {\it global} SO(3,1) group.
In fact the TEGR was previously considered in ref. \cite{Maluf1}
with a local SO(3,1) symmetry. In order to make clear this point
let us recall that relation (1) can be written in terms of
$e^a\,_\mu$ and an arbitrary spin connection $\omega_{\mu ab}$.
As discussed in \cite{Maluf1}, $e^a\,_\mu$ and $\omega_{\mu ab}$
are not related to each other via field equations.
This arbitrary connection can be {\it identically} written as
$\omega_{\mu ab}=\,^0\omega_{\mu ab} + K_{\mu ab}$, where
$K_{\mu ab}$ is the same as above but now
the torsion tensor is given by
$T^a\,_{\mu \nu}=\partial_\mu e^a\,_\nu-\partial_\nu e^a\,_\mu+
\omega_\mu\,^a\,_b\, e^b\,_\nu-\omega_\nu\,^a\,_b\, e^b\,_\mu$.
Substituting $\omega_{\mu ab}$ into the scalar curvature
$R(e,\omega)=e^{a\mu}e^{b\nu}R_{ab\mu \nu}(\omega)$ we obtain
the identity

$$eR(e,\omega)=eR(e)+e({1\over 4}T^{abc}T_{abc}+
{1\over 2}T^{abc}T_{bac}-T^aT_a)-2\partial_\mu(eT^\mu)\;.$$

\noindent The teleparallel space is determined by the vanishing
of $R(e,\omega)$.

It was shown in ref.\cite{Maluf1} that
the Hamiltonian formulation of 
the TEGR with a {\it local} SO(3,1) symmetry cannot be made
consistent since the constraint algebra does not ``close", and
therefore the dynamical evolution of the field quantities is not
well defined. A well established Hamiltonian formulation can only
be achieved if the SO(3,1) is turned into a global symmetry group.
The requirement of the vanishing of the curvature tensor
$R^a\,_{b\mu \nu}(\omega)$ that appears on the left hand side of
the identity above has the ultimate effect of discarding the
connection  $\omega_{\mu ab}$. A global SO(3,1) symmetry
leads to a theory with well defined initial value problem.
Therefore we can dispense with the local symmetry of the theory
together with the constraint of vanishing curvature, both
considered previously in \cite{Maluf1}. Kopczy\'nski
(ref.\cite{Kop2}, section 4) has argued that even at the Lagrangian
level one can require $\omega_{\mu ab}=0$ and obtain the same
field equations. As a matter of fact the present results confirm
this point of view.\par

Thus the Hamiltonian will display a global SO(3) symmetry.
We remark that the local SO(3) symmetry group has recently
played a special role in connection with the Hamiltonian formulation
of gravity theories, as developed in \cite{Lopez}. In the latter
it is carried out a Poincar\'e invariant foliation of the
space-time in which the SO(3) group is taken as the classification
subgroup of the Poincar\'e group (rather than the Lorentz group).
A Hamiltonian formalism based on a nonlinear realization of the
Poincar\'e group is constructed and applied to the Einstein-Cartan
theory.

\bigskip

The major motivation for considering the TEGR resides in the fact
that it is possible to
make definite statements about the energy and
momentum of the gravitational field. In the 3+1 formulation of the
TEGR we find that  the Hamiltonian and vector constraints contain
each one a divergence in the form of a scalar and vector densities,
respectively, that can be identified as the energy and momentum
{\it densities} of the gravitational field\cite{Maluf3}.
Therefore the Hamiltonian and vector constraints are considered
as energy-momentum equations. This identification has proven to
be consistent, and has shown that the TEGR provides a natural
setting for investigations of the gravitational energy. Several
relevant applications have been presented in the literature.
Among the latter we point out investigations on the gravitational
energy of rotating black holes\cite{Maluf4} (the evaluation
of the irreducible mass of a Kerr black hole) and
of Bondi's radiating metric\cite{Maluf5}.\par
\bigskip
In this paper we carry out the Hamiltonian formulation of an
arbitrary teleparallel theory, quadratic in the torsion tensor
just like in (2). The Lagrangian density to be considered
describes a three-parameter family of teleparallel theories.
We want to investigate the existence of theories 
that satisfy the only criterium of having a well
defined Hamiltonian formulation, which amounts to having a well
posed initial value problem. For this purpose we adopt
the field quantity definitions of Hayashi and
Shirafuji\cite{Shirafuji}. We also make use of their
analysis of the Newtonian limit of these theories and of the
restrictions implied by this requirement.

The investigation will
be carried out along the lines of ref. \cite{Maluf1}. As in
the latter, we will impose the time gauge condition to the
tetrad field (this condition is important in order to establish
a comparison with \cite{Maluf1}). A consistent implementation of
the Legendre transform reduces the three-parameter to a
one-parameter family of theories. The latter constitutes a well
defined theory with only first class
constraints. The free parameter is fixed by requiring the
gravitational field to exhibit the Newtonian limit.
The resulting theory is just the TEGR.

In section II we establish the definitions and present
the Lagrangian formulation of the teleparallel theory. The
Hamiltonian formulation is established in section III.
The relevant details of the Legendre transform will be
presented in this section. In the last section we present
our final comments.\par
\bigskip
\noindent Notation: spacetime indices $\mu, \nu, ...$ and SO(3,1) 
indices $a, b, ...$ run from 0 to 3. In the 3+1 decomposition latin 
indices from the middle of the alphabet indicate space indices according 
to $\mu=0,i,\;\;a=(0),(i)$.
The flat spacetime metric is fixed by $\eta_{(0)(0)}=-1$.\par

\bigskip
\bigskip
\noindent {\bf II. The Lagrangian formulation of an
arbitrary teleparallel theory}\par
\bigskip
We begin by presenting the four basic postulates that the Lagrangian
density for the gravitational field in empty space-time, in the
teleparallel geometry,  must satisfy.
It must be (i) invariant under general coordinate transformations,
(ii) invariant under global SO(3,1) transformations, (iii) invariant
under parity transformations and (iv) quadratic in the torsion tensor.
The most general Lagrangian density can be written as

$$L_0\;=\;-ke(c_1 t^{abc}t_{abc}+
c_2v^a v_a + c_3a^b a_b)\;,\eqno(4)$$

\noindent where $c_1, c_2, c_3$ are constants and

$$t_{abc}\;=\;{1\over 2}(T_{abc}+T_{bac})+
{1\over 6}(\eta_{ac}v_b+\eta_{bc}v_a)-
{1\over 3} \eta_{ab}v_c\;,\eqno(5.1)$$

$$v_a\;=\;T^b\,_{ba}\;=\;T_a\;,\eqno(5.2)$$

$$a_a\;=\;{1\over 6}\varepsilon_{abcd}T^{bcd}\;,\eqno(5.3)$$

$$T_{abc}\;=\;e_b\,^\mu e_c\,^\nu T_{a\mu \nu}\;.$$

\noindent Definitions (5)
correspond to the irreducible components of the torsion tensor.
As we mentioned earlier, we are departing from
Hayashi and Shirafuji's notation\cite{Shirafuji}. Our analysis will
make contact both with Ref. \cite{Maluf1} and with
Ref. \cite{Shirafuji}. We are considering an {\it extended}
teleparallel theory in the sense of M\"uller-Hoissen and
Nitsch\cite{Muller}.

In order to carry out the Hamiltonian formulation in the next
section we need to rewrite the three terms of $L_0$ in order
to make explict the appearance of the torsion tensor. Therefore
we rewrite $L_0$ as

$$L_0\;=\;-ke(c_1X^{abc}T_{abc}+c_2Y^{abc}T_{abc}+
c_3Z^{abc}T_{abc})\;,\eqno(6)$$

\noindent with the following definitions:

$$X^{abc}\;=\;{1\over 2}T^{abc}+{1\over 4}T^{bac}-{1\over 4}T^{cab}
+{1\over 4}(\eta^{ac}v^b-\eta^{ab}v^c)\;,\eqno(7.1)$$

$$Y^{abc}\;=\;{1\over 2}(\eta^{ab}v^c-\eta^{ac}v^b)\;,\eqno(7.2)$$

$$Z^{abc}\;=\;-{1\over 18}(T^{abc}+T^{bca}+T^{cab})\;.\eqno(7.3)$$

\noindent The definitions above satisfy $X^{abc}=-X^{acb}$,
$Y^{abc}=-Y^{acb}$ and $Z^{abc}=-Z^{acb}$. $X^{abc}, Y^{abc}$ and
$Z^{abc}$ have altogether the same number of independent components
of $T^{abc}$. It is not difficult to verify that
$X^{abc}+X^{bca}+X^{cab}\equiv 0$.

Let us define the field quantity $\Sigma^{abc}$ by

$$\Sigma^{abc}\;=\;c_1X^{abc}+c_2Y^{abc}+c_3Z^{abc}\;,\eqno(8)$$

\noindent which allow us to further rewrite $L_0$ according to the
notation of Ref. \cite{Maluf1}:

$$L_0\;=\;-ke\,\Sigma^{abc}T_{abc}\;.\eqno(9)$$

\noindent We note that if the constants $c_i$ satisfy

$$c_1={2\over 3}\;,\;\;\;c_2=-{2\over 3}\;,\;\;\;
c_3={3\over 2}\;,\eqno(10)$$

\noindent then $\Sigma^{abc}$ reduces to the corresponding quantity
of the TEGR \cite{Maluf1}:

$$\Sigma^{abc}_{_{_{TEGR}}}\;=\;{1\over 4}(T^{abc}+T^{bac}-T^{cab})+
{1\over 2}(\eta^{ac}v^b-\eta^{ab}v^c)\;,\eqno(11)$$

\noindent for which we have

$$\Sigma^{abc}_{_{_{TEGR}}}T_{abc}\;=\;
{1\over 4}T^{abc}T_{abc} + {1\over 2}T^{abc}T_{bac}-T^aT_a\;.
\eqno(12)$$

\bigskip

In order to carry out the 3+1 decomposition of the theory we need
a first order differential Lagrangian density. It will be achieved
through the introduction of an auxiliary field quantity
$\Delta_{abc}$, according to the procedure developed in
\cite{Maluf1}. Thus we consider the Lagrangian density

$$L(e,\Delta)\;=\;-ke(c_1\Theta^{abc}+c_2\Omega^{abc}+
c_3\Gamma^{abc})(\Delta_{abc}-2T_{abc})\;,\eqno(13)$$

\noindent where $\Theta^{abc}$, $\Omega^{abc}$ and $\Gamma^{abc}$
are defined in similarity with $X^{abc}$, $Y^{abc}$ and $Z^{abc}$,
respectively:

$$\Theta^{abc}={1\over 2}\Delta^{abc}+{1\over 4}\Delta^{bac}-
{1\over 4}\Delta^{cab}+
{1\over 4}(\eta^{ac}\Delta^b-\eta^{ab}\Delta^c)\;,\eqno(14.1)$$

$$\Omega^{abc}=
{1\over 2}(\eta^{ab}\Delta^c-\eta^{ac}\Delta^b)\;,\eqno(14.2)$$

$$\Gamma^{abc}=-{1\over {18}}(\Delta^{abc}+\Delta^{bca}+
\Delta^{cab})\;.\eqno(14.3)$$

\noindent The three quantities above are anti-symmetric in the
last two indices.

The field equations are most easily obtained by making use of
the three identities satisfied by these expressions:

$$X^{abc}\Delta_{abc}\;=\;\Theta^{abc}T_{abc}\;,\eqno(15.1)$$

$$Y^{abc}\Delta_{abc}\;=\;\Omega^{abc}T_{abc}\;,\eqno(15.2)$$

$$Z^{abc}\Delta_{abc}\;=\;\Gamma^{abc}T_{abc}\;.\eqno(15.3)$$

\noindent These identities turn out to be useful in the variation
of the action integral. We note in addition that since
$\Theta^{abc}\Delta_{abc}$ is quadratic in $\Delta_{abc}$ it
follows that

$$\delta(\Theta^{abc}\Delta_{abc})\;=
\;2\Theta^{abc}\delta(\Delta_{abc})\;,\eqno(16)$$

\noindent and similarly for $\Omega^{abc}$ and $\Gamma^{abc}$
(this result can be verified by explicit calculations).
Because of (16) the variation of
$-ke\,c_1\Theta^{abc}(\Delta_{abc}-2T_{abc})$ with respect to
$\Delta_{abc}$ is given by

$$\delta \lbrace -ke\,c_1\Theta^{abc}(\Delta_{abc}-2T_{abc})\rbrace
=-2ke\,c_1(\Theta^{abc}-X^{abc})\delta(\Delta_{abc})\;,\eqno(17)$$

\noindent and likewise for the other terms in $L$.
Therefore the field equations arising from (13) with
respect to variations of $\Delta_{abc}$ are given by

$$c_1(\Theta^{abc}-X^{abc})+c_2(\Omega^{abc}-Y^{abc})
+c_3(\Gamma^{abc}-Z^{abc})\;=\;0\;.\eqno(18)$$

\noindent The only solution of (18) for arbitrary constants $c_i$
is given by

$$\Delta_{abc}\;=\;T_{abc}\;=
\;e_b\,^\mu e_c\,^\nu T_{a\mu\nu}\;.\eqno(19)$$

\noindent Note that (18) represents 24 equations for 24 unknown
quantities $\Delta_{abc}$.\par
\bigskip

If the constants $c_i$ satisfy (10) then the field
equations obtained with respect to variations of $e^{a\mu}$
are, in view of (3), precisely equivalent to Einstein's
equations.

It should me mentioned that teleparallel theories also arise as
effective theories in the context of Poincar\'e gauge theories of
gravity by means of a modified double duality ansatz
(see, for instance, Baekler et. al.\cite{Baekler1}), however in
the limit of vanishing curvature of the Riemann-Cartan manifold.\par
\bigskip
\bigskip
\bigskip
\noindent {\bf III. The Hamiltonian formulation}\par
\bigskip
Although in this section we still maintain the notation
given at the end of section I, 
we will make a change of notation regarding the tetrad
field $e^a\,_\mu$. The space-time tetrad field considered
in the last section will be denoted here as
$^4e^a\,_\mu$, to emphasize  that it is the tetrad field of the
four-dimensional space-time. In a 3+1 decomposition 
the space-time tetrad field does not coincide with the tetrad
field restricted (projected) to the three-dimensional spacelike
hypersurface.

We adopt the standard 3+1 decomposition for the tetrad field:

$$^4e^a\,_k=e^a\,_k\;,$$

$$^4e^{ai}=e^{ai}+{{N^i}\over N}\eta^a\;,$$

$$e^{ai}=\bar g^{ik}e^a\,_k\;\;,\;\;\;\;\;\eta^a=-N\,^4e^{a0}\;,$$

$$^4e^a\,_0=N^i e^a\,_i+N\eta^a\;,$$

$$^4e=Ne=N\sqrt{e^a\,_i e_{aj}}\;,\eqno(20)$$

\noindent where $g_{ij}=e^a\,_i e_{aj}$ and $\bar g^{ij}g_{jk}=
\delta^i_k$. The vector $\eta^a$ satisfies

$$\eta_ae^a\,_k=0\;\;,\,\,\,\,\eta_a \eta^a=-1\;.$$

\noindent It follows that

$$e^{bk}e_{bj}=\delta^k_j\;,$$

$$e^a\,_ie^{bi}=\eta^{ab}+\eta^a \eta^b\;.$$

\noindent The components $e^{ai}$ and $e^a\,_k$ are now
restricted to the spacelike hypersurface.

The Hamiltonian formulation will be established by rewriting the
Lagrangian density (13) in the form $L=p\dot q -H$. There is no
time derivative of $^4e_{a0}$, and therefore we will enforce the
corresponding momentum $P^{a0}$ to vanish from the outset.

In analogy with (8) let us define the quantity
$\Lambda^{abc}$,

$$\Lambda^{abc}=c_1\Theta^{abc}+c_2\Omega^{abc}+
c_3\Gamma^{abc}\;,\eqno(21)$$

\noindent in terms of which we define $P^{ai}$,
the momentum canonically conjugated to $e_{ai}$:

$$P^{ai}\;=\;4k\;^4e\,\Lambda^{a0i}\;=
\;4ke\,e_b\,^i\eta_c\Lambda^{abc}\;.\eqno(22)$$

In a first step the Lagrangian density is written as

$$L\;=\;P^{ai}\dot e_{ai}+\,^4e_{a0}\partial_i P^{ai}
+2Nke\,\Lambda^{aij}T_{aij}+N^k P^{ai}T_{aik}$$

$$-Nke\Lambda^{abc}\Delta_{abc}-
\partial_i(P^{ai}\,^4e_{a0})\;,\eqno(23)$$

\noindent where $\Lambda^{aij}=e_b\,^i e_c\,^j\Lambda^{abc}$.
The task of writing $L$ in terms of $e_{ai}$,
$P^{ai}$ and Lagrange multipliers is not trivial.
The troublesome term is  $-Nke\Lambda^{abc}\Delta_{abc}$.
We will make use of the field equations (19) and identify

$$\Delta _{a\mu\nu}=T_{a\mu\nu}$$

\noindent in $L$. The Legendre transform would be
straightforward if, in view of (22), $\Lambda^{aij}$ would
depend only on $e_{(i)j}$ and its spatial  derivatives,
which at this point is not the case. The Hamiltonian density
cannot depend on the components $\Delta_{a0j}=T_{a0j}$,
associated to the velocities $\dot e_{ai}$. Therefore
these components will have to be eliminated
in the Legendre transform. We find it convenient to
establish a decomposition for $\Lambda^{abc}$ in order to
distinguish the components that  contribute to the canonical
momentum $P^{ai}$. It is given by

$$\Lambda^{abc}={1\over {4ke}}(\eta^b e^c\,_i P^{ai}-
\eta^c e^b\,_i P^{ai})+e^b\,_i e^c\,_j\Lambda^{aij}\;.\eqno(24)$$

\noindent  The quantity  $\Lambda^{aij}$ in the expression above
contains ``velocity"  terms $\Delta_{a0j}$ that cannot be
inverted and written in terms of $P^{ai}$.
However these terms will not be
present in final expression of $L$. This feature will be achieved
in view of the Schwinger's time gauge condition\cite{Schwinger}

$$\eta^a\;=\;\delta^a_{(0)}\;,\eqno(25)$$

\noindent that implies $^4e_{(k)}\,^0=e^{(0)}\,_i=0$. (25)  is
assumed to hold from now on, i.e., it is assumed to hold
before varying the action. As a consequence $\dot e_{(0)i}=0$.
Taking into account definitions (21) and (22) we find by explict
calculations that

$$P^{(0)k}=-2ke(c_1+c_2)T^{(0)}\,_{(0)}\,^k
+ke(c_1-2c_2)T^k\;.\eqno(26)$$

\noindent We will soon return to this expression. The time gauge
condition actually reduces the configuration space of the
theory, and also reduces the symmetry group from the
SO(3,1) to the global SO(3) group. As a consequence the
teleparallel geometry is restricted to the three-dimensional
spacelike hypersurface.

In the following we will rewrite the various components of $L$
in terms of canonical quantities. First, it is not difficult
to verify that

$$^4e_{a0}\partial_i P^{ai}=N^ke_{(j)k}\partial_i P^{(j)i}-
N\partial_i P^{(0)i}\;.\eqno(27)$$

\noindent We also have

$$-\partial_i (\,^4e_{a0}P^{ai})=
\partial_i(NP^{(0)i})-
\partial_i(N_kP^{ki})\;.\eqno(28)$$

\bigskip
\bigskip

Next we consider
$-Nke\Lambda^{abc}\Delta_{abc}$. By making use of (20), (24) and
(25) it can be rewritten as

$$-Nke\Lambda^{abc}\Delta_{abc}=N\biggl( {{c_1}\over 4}+
{{c_3}\over{18}}\biggr)^{-1}\lbrace  {1\over{16ke}}(P^{ij}P_{ji}-
P^{(0)i}P^{(0)}\,_i)$$

$$+{1\over 2}e_{(m)i}P^{(m)}\,_j\Lambda^{(0)ij}-
kee^{(m)}\,_ie_{(n)}\,^k\Lambda^{(n)ij}\Lambda_{(m)kj}\rbrace$$

$$+N\biggl( {{c_1}\over {2c_2}} -1\biggr)\lbrace {1\over {48ke}}
(P^2-P^{(0)i}P^{(0)}\,_i)+
{1\over 6}P^{(0)}\,_ke_{(m)j}\Lambda^{(m)jk}-
{1\over3}kee_{(m)i}\Lambda^{(m)ik}e_{(n)j}\Lambda^{(n)j}\,_k\rbrace$$

$$+N\biggl( {{c_3}\over 9}-{{c_1}\over 4}\biggr)
\lbrace {1\over 4}\Delta_{ij(0)}P^{ij}-
ke\Delta_{i(0)j}\Lambda^{(0)ij}-ke\Delta_{ikj}\Lambda^{kij}
-{1\over 4}\Delta_{(0)(0)i}P^{(0)i}-
{1\over 4}\Delta_{(0)ij}P^{ij}\rbrace\;.\eqno(29)$$

\noindent Spatial indices are raised
and lowered with the help of $e_{(i)j}$ and $e^{(k)m}$.

We note that the first term on the second line of the
expression above actually reads (except for the lapse function
and for the multiplicative term)

$${1\over 2} P_{\lbrack ij \rbrack}\Lambda^{(0)ij}\;,$$

\noindent where $\lbrack .. \rbrack$ denotes antisymmetrization.
The expression of $\Lambda^{(0)ij}$ contains ``velocity"
terms $\Delta_{a0j}$. We know, however, that in tetrad
type theories of gravity the anti-symmetric part of the momentum
is contrained to vanish. Let us evaluate the expression of
$P_{\lbrack i j \rbrack }$ directly from its definition (22), taking
into account expressions (19) and (21) for $\Lambda^{abc}$. It is
given by

$$P_{\lbrack ij \rbrack }+ke\biggl(c_1+{2\over 9}c_3\biggr)T_{(0)ij}
+ke\biggl(c_1-{4\over 9}c_3\biggr)T_{\lbrack i\vert(0)\vert j\rbrack}
=0\;.\eqno(30)$$

\noindent In the time gauge the term
$T_{(0)ij}=\partial_i e_{(0)j}-\partial_j e_{(0)i}$ vanishes.
We will return to this expression latter together with
expression (26).\par
\bigskip
\bigskip
The third term to be considered in $L$ is $N^kP^{ai}T_{aik}$.
In view of the time gauge condition it reads

$$N^kP^{ai}T_{aik}=N^kP^{(i)j}T_{(i)jk}\;.\eqno(32)$$

\bigskip
\bigskip
Lastly, we work out the remaining term
$2Nke\Lambda^{aij}T_{aij}$. The time gauge condition simplifies
this expression in two respects. First, 
the term for which $a=(0)$ vanishes. Thus
$2Nke\Lambda^{aij}T_{aij}=2Nke\Lambda^{kij}T_{kij}$. Second,
because of (25) it can be shown by explicit calculations that
$\Lambda^{kij}T_{kij}$ does not contain terms of the type
$\Delta_{a0j}=T_{a0j}$. Therefore $\Lambda^{kij}T_{kij}$ is totally
projected in the spacelike hypersurface. \par
\bigskip
\bigskip
We are now in a position of bringing back
expressions (26)-(30) to the
Lagrangian density (23). Before carrying out the substitution
we can establish the conditions under which the Lagrangian
density will be exempt of the terms $T_{a0j}$. From expression
(26) we observe that we must demand

$$c_1+c_2=0\;.\eqno(33)$$

\noindent Next we see that the last line of (29), which contains
several $\Delta_{a0j}$ type terms, is discarded if we require

$$c_1-{4\over 9}c_3=0\;.\eqno(34)$$

\noindent We observe from (30) that (34) ensures that
$P_{\lbrack ij \rbrack}$ vanishes in the time gauge,

$$P_{\lbrack ij \rbrack}=0\;,\eqno(35)$$

\noindent which in turn makes (29) exempt of the term
${1\over 2} P_{\lbrack ij \rbrack}\Lambda^{(0)ij}$. Thus
$P_{\lbrack ij \rbrack}$ will enter the Hamiltonian density
$H=p\dot q-L$ multiplied by a Lagrange multiplier.

We can finally provide the ultimate expression of $L$ by
collecting terms that multiply the lapse and shift functions.
We choose to write it in terms of the constant $c_1$.
Note that $\Lambda^{kij}$ can now be substituted by
$\Sigma^{kij}$, which is a function of $e_{(i)j}$ {\it only}.
The final expression reads

$$L\;=\;P^{(i)j}\dot e_{(i)j}+
NC+N^kC_k+\lambda^{ij}P_{\lbrack ij \rbrack}
-\partial_i(3c_1\,ke T^i+N_kP^{ki})\;,\eqno(36)$$

\noindent where $\lbrace \lambda^{ij} \rbrace$ are Lagrange
multipliers. The Lagrangian density above is invariant under
the global SO(3) group. The Hamiltonian and vector
constraints are given by

$$C={1\over {6c_1\,ke}}\biggl(P^{ij}P_{ji}-{1\over 2}P^2\biggr)
+ke\Sigma^{kij}T_{kij}-\partial_i(3c_1\,ke T^i)\;,\eqno(37)$$

and

$$C_k=e_{(j)k}\partial_i P^{(j)i}+
P^{(j)i}T_{(j)ik}\;,\eqno(38)$$

\noindent where $T^i=\bar g^{ik}T_k=
\bar g^{ik}e^{(m)j}T_{(m)jk}$; $\Sigma^{kij}$ is defined by
(8) together with conditions (33) and (34).
These latter conditions reduce the theory defined by (4)
to a one-parameter theory.\par
\bigskip
\bigskip

\noindent {\bf IV. Discussion}\par
\bigskip
We observed that the Legendre transform has reduced the
three-parameter to a one-parameter family of teleparallel
theories. A consistent implementation of the Legendre
transform is a necessary condition for the Hamiltonian
formulation, but not sufficient. The
complete canonical formulation demands further crucial
investigations. It is also necessary to verify whether the
constraints constitute a first class set, namely, whether the
algebra of constraints ``closes". In the TEGR the Hamiltonian
formulation and the constraint algebra have been obtained
in \cite{Maluf1}. The Hamiltonian constraint of the latter is
very similar to (37), except for the presence of $c_1$ in the
three terms of $C$. By making $c_1={2\over 3}$ expression
(37) becomes precisely the Hamiltonian constraint of
\cite{Maluf1}. Constraint (38) is the same as in \cite{Maluf1}.

Let us write $\Sigma^{kij}$ that appears in (37) in terms of
$c_1$, using definition (8) and conditions (33) and (34). It
reads

$$\Sigma^{kij}={{3c_1}\over 2}\biggl(
{2\over 3} X^{kij}-{2\over 3}Y^{kij}+{3\over 2}Z^{kij}\biggr)=
{{3c_1}\over 2}\,\Sigma^{kij}_{_{_{TEGR}}}\;.\eqno(39)$$

\noindent where $\Sigma^{kij}_{_{_{TEGR}}}$ is restricted to the
three-dimensional spacelike hypersurface and is defined in
similarity with (11). By means of conditions (33) and (34)
Lagrangian density (9) becomes

$$L_0=-{{3c_1}\over2}k\,e\,
\Sigma^{abc}_{_{_{TEGR}}}T_{abc}\;.\eqno(40)$$

\noindent We observe then that by defining

$$k'={{3c_1}\over 2}\,k\;,$$

\noindent we can rewrite (37) according to

$$C={1\over {4k'e}}\biggl(P^{ij}P_{ji}-{1\over 2}P^2\biggr)
+k'e\Sigma^{kij}_{_{_{TEGR}}}T_{kij}-
\partial_i(2k'eT^i)\;.\eqno(41)$$

\noindent Except for $k'$ expression above is {\it exactly} the
Hamiltonian constraint of ref. \cite{Maluf1}. The constant
$k'$ does {\it not} affect the evaluation of Poisson brackets
between (38), (41) and $P^{\lbrack ij \rbrack}$.
Thus we conclude that the constraint algebra determined by
(41) and (38) is exactly the same of ref. \cite{Maluf1}.
Therefore (37) (or (41)) and (38) are first class
constraints. As a consequence field quantities have a well
defined time evolution. The fixing of $k'$ is  related
to the Newtonian behaviour of the gravitational field.

Hayashi and Shirafuji\cite{Shirafuji} have analyzed the
Lagrangian field equations derived from (4). In particular they
have investigated the conditions under which a correct
Newtonian approximation is obtained by studying solutions of the
field equations that yield static and isotropic gravitational
fields. Without imposing any {\it a priori} restriction on the
parameters $c_i$ they concluded that the
Newtonian limit is verified for a class of solutions provided

$$c_2=-{{(c_1-{2\over 3})}\over {(1-{9\over 8}c_1)}}-
{2\over 3}\;.$$

\noindent No condition fixes $c_3$. By imposing the mandatory
condition (33), $c_1+c_2=0$, in the expression above it
follows that $c_2=-{2\over 3}$ and $c_1={2\over 3}$.
Hence we finally arrive at the TEGR.

Lenzen\cite{Lenzen} and Baekler et. al.\cite{Baekler2} have
shown the emergence of free functions in exact torsion solutions
in Poincar\'e gauge theories (PGT) of gravity. Therefore it is
worth examining this  question here. The Hamiltonian formulation
developed above provides a suitable framework for such analysis.
The emergence of free functions is related to the selection
of appropriate triads (tetrads) for the space (space-time).
The counting of degrees of freedom here is the same as in the usual
ADM\cite{ADM} (metrical) formulation, except that there are 9
triad components in (36) rather than 6 metric functions as in the 
ADM  formulation
(the imposition of $P_{\lbrack ij \rbrack}=0$ together with the
field equation $\dot e_{(i)j}(x)=\lbrace e_{(i)j}(x),H\rbrace$
leads to an expression for $\lambda_{ij}$ in terms of $e_{(i)j}$).
Therefore there are 3 extra (undetermined) triad components.
However we have discussed in \cite{Maluf5} that these 3 components
may be  fixed in the context of isolated material systems by
the asymptotic behaviour in the limit
$r \rightarrow \infty$,

$$e_{(i)j} \approx \eta_{ij}+
{1\over 2}h_{ij}({1\over r})\;,\eqno(42)$$

\noindent where $h_{ij}=h_{ji}$ is the first space dependent term
in the asymptotic expansion of the metric tensor $g_{ij}$. It is
not difficult to verify that this condition fixes {\it uniquely}
a triad to a three-dimensional metric tensor. In fact this condition
has already been suggested by M\o ller\cite{Moller1,Moller2}
for the same purpose.
It turns out that in the TEGR condition (42) is essential in order
obtain the ADM energy out of the scalar density
$\partial_i(eT^i)$ (with appropriate multiplicative constants)
in the Hamiltonian constraint\cite{Maluf3}.
A further condition on the triads is also essential in
the TEGR, mainly in respect to the definition of gravitational
energy: we must have $T_{(i)jk}=0$ if we make the physical
parameters of the metric tensor (such as mass, angular momentum, etc)
vanish. Triads that lead to a vanishing torsion tensor in any
coordinate system are called {\it reference space}
triads\cite{Maluf4} (all applications of the
TEGR\cite{Maluf2,Maluf3,Maluf4,Maluf5} have taken into account the
reference space triads). 
The two conditions above associate uniquely
a metric tensor to  triads (tetrads) components, make the latter
exempt of free functions and lead to a well defined notion of
gravitational energy.

Baekler and Mielke\cite{Baekler2} consider the Hamiltonian
formulation of the most general class
of PGT theories, constructed out of
tetrad fields $e_{a\mu}$ and connections $\omega_{\mu ab}$. They
claim that a first class algebra is achieved irrespective of any
prior gauge condition on $\omega_{0ab}$, and for a theory with
arbitrary multiplicative constants for the squared torsion and
curvature terms (the $a_i$ constants of the PGT theories are
related to $c_i$ according to $c_1=-{2\over 3}a_1$,
$c_2=-{1\over 3}a_2$ and $c_3=3a_3$).
In view of this result the time evolution of
field quantities would be, in principle, well defined
(Hecht et. al.\cite{Hecht} also consider the initial value problem
for some PGT theories; they argue that there is freedom in the
choice of the PGT parameters in the Lagrangian density such that
the theory acquires a mathematically well defined initial value
problem, but no Hamiltonian analysis is carried out).
Certainly the analysis of \cite{Baekler2} is not in  agreeement with
that of ref.\cite{Maluf1}, where the fixation of $\omega_{0ab}$ is
mandatory. However it was shown by Kopczy\'nski\cite{Kop1} and
M\"uller-Hoissen and Nitsch\cite{Muller} that the TEGR defined
in terms of tetrad fields and connections $\omega_{\mu ab}$,
supplemented by the condition of vanishing curvature
(as developed in \cite{Maluf1}) faces
difficulties with respect to the Cauchy problem. They have shown
that in general six components of the torsion tensor are not
determined from evolution of the initial data. On the other hand
in the context of \cite{Maluf1}
the constraints of the theory become a first class set provided
we fix the six quantities $\omega_{0ab}=0$ before varying the
action (this point is also discussed in \cite{Maluf5}). Although
we have no proof we believe that the two properties above
(the failure of the Cauchy problem and the fixation of
$\omega_{0ab}$) are related to each other. For this reason we
dispense with the constraint of vanishing curvature and consider
the theory defined by (2).

We note finally that a similar analysis has been developed by
Blagojevi\'c and Nikoli\'c\cite{BN}, who considered
PGT theories of the type $R+R^2+T^2$.
Although the latter has been worked
out in the framework of Riemannian geometry, with both
tetrad and connection fields, it would be
possible, in principle, to establish a comparison of their work
with our analysis. However, the constraints of Ref. \cite{BN}
are only formally indicated. They are not explicitly
expressed in terms of canonical variables as in
(37) and (38), and therefore an objective comparison cannot
be made. Moreover the constraint algebra in the particular case
$c_1={2\over 3}, c_2=-{2\over 3}$, $c_3={3\over 2}$,
as obtained in ref.\cite{Maluf1}, has not been established in
this earlier investigation. A recent investigation\cite{Yo} on
the Hamiltonian formulation of PGT theories
has been carried out along the lines of \cite{BN}. Instead of
actually performing an ordinary Legendre transform the authors
make use of the {\it if constraint} formalism of \cite{BN}.
For the restricted sector of torsion squared terms they obtain
a {\it three} parameter Hamiltonian density  with second class
constraints. However their Hamiltonian analysis does not single
out the ``viable" conditions $a_1+2a_3=0$ and $2a_1+a_2=0$
(in the notation of \cite{Yo}), which they take into account in
order to carry out their analysis. These conditions 
are enforced by hand and  correspond precisely to
conditions (33) and (34).\par

\end{document}